\documentclass[aps,showpacs,prb,reprint,longbibliography,superscriptaddress]{revtex4-2}

\usepackage[colorlinks=true,linkcolor=blue,citecolor=blue,urlcolor=blue]{hyperref}
\usepackage{setspace} 
\usepackage{graphicx}
\usepackage{amsmath}
\usepackage{color}
\usepackage{amsmath}
\usepackage{mathtools}
\usepackage{amssymb}
\usepackage{verbatim}
\usepackage{physics}
\usepackage{latexsym}
\usepackage{enumerate} 
\usepackage{bm} 
\usepackage[caption=false]{subfig}
\usepackage{multirow}
\usepackage{booktabs}
\usepackage{siunitx}

\usepackage{float}

\usepackage{lipsum}

\begin{document}

\title{\textbf{\Large{Phonon-mediated electron attraction in SrTiO$_3$ via the generalized Fr\"ohlich and deformation potential mechanisms}}}

\author{Christopher J. N. Coveney}
\affiliation{Department of Physics, University of Oxford, Oxford OX1 3PJ, United Kingdom}
\author{Norm M. Tubman}
\email{norman.m.tubman@nasa.gov}
\affiliation{NASA Ames Research Center, Moffett Field, California 94035, United States}
\author{Chih-En Hsu}
\affiliation{Department of Physics, Tamkang University, New Taipei City 251301, Taiwan}
\affiliation{Mork Family Department of Chemical Engineering and Materials Science, University of Southern California, Los Angeles, California 90089, USA}
\author{Andres Montoya-Castillo}
\affiliation{Department of Chemistry, University of Colorado Boulder, Boulder, CO 80309, USA}
\author{Marina R. Filip}
\affiliation{Department of Physics, University of Oxford, Oxford OX1 3PJ, United Kingdom}
\author{Jeffrey B. Neaton}
\affiliation{Materials Sciences Division, Lawrence Berkeley National Laboratory, Berkeley, California 94720, United States}
\affiliation{Department of Physics, University of California Berkeley, Berkeley, California 94720, United States}
\affiliation{Kavli Energy NanoScience Institute at Berkeley, Berkeley, California 94720, United States}
\author{Zhenglu Li}
\affiliation{Mork Family Department of Chemical Engineering and Materials Science, University of Southern California, Los Angeles, California 90089, USA}
\author{Vojtech Vlcek}
\affiliation{Department of Chemistry and Biochemistry, University of California, Santa Barbara, CA 93106, USA}
\affiliation{Materials Department, University of California, Santa Barbara, CA 93106-9510, USA}
\author{Antonios M. Alvertis}
\email{antoniosmarkos.alvertis@nasa.gov}
\affiliation{KBR, Inc., NASA Ames Research Center, Moffett Field, California 94035, United States}

\date{\today}
\begin{abstract}
Superconductivity in doped SrTiO$_3$ was discovered in 1964, the first superconducting transition observed in a doped semiconductor. However, the mechanism of electron pairing in SrTiO$_3$ 
remains a subject of debate. By developing a theoretical framework to incorporate dynamical lattice screening
in the electronic Coulomb interactions of semiconductors
and insulators, we demonstrate analytically that linear
long-range coupling of electrons to multiple longitudinal optical phonons, 
described by
a generalized Fr\"ohlich mechanism, can result
in superconductivity in SrTiO$_3$. Moreover, by combining our theory with first-principles calculations, we 
reveal
an additional attractive interaction between
electrons in SrTiO$_3$ due to the deformation potential mechanism, arising from the mixed ionic-covalent character of the Ti-O bond. Our results
may have implications for the emergence of phonon-mediated electron attraction and superconductivity in a broad range of materials.
\end{abstract}

\maketitle

The prediction that doped semiconductors can exhibit phonon-mediated superconductivity was made in 1964 in the seminal works by Cohen~\cite{PhysRev.134.A511,RevModPhys.36.240}, leading to
the discovery of superconductivity in doped $\text{SrTiO}_3$ in the same year~\cite{PhysRevLett.12.474}. Since then, superconductivity has been discovered in several semiconductors, including GeTe~\cite{PhysRevLett.124.047002,Cheng2024}, SnTe~\cite{PhysRev.177.704,PhysRevB.88.140502}, and MoS$_2$~\cite{doi:10.1126/science.aab2277}. Moreover, interfaces of $\text{SrTiO}_3$ with LaAlO$_3$ and LaTiO$_3$ exhibit superconductivity~\cite{Gariglio_2009,Biscaras2010,Tan2013}, generating
interest in interface engineering to design high-temperature
superconductors. Therefore,  
understanding the mechanism of $\text{SrTiO}_3$ superconductivity is key towards designing new types of superconductors. 

To date, the precise mechanism of the pairing between electrons in $\text{SrTiO}_3$ remains elusive. 
As $\text{SrTiO}_3$ exhibits strong electron-phonon coupling~\cite{PhysRevLett.121.226603,PhysRevResearch.2.043296}, several works have
investigated the possibility of phonon-mediated superconductivity~\cite{PhysRev.134.A511,PhysRev.163.380,doi:10.1073/pnas.1604145113,Gorkov2017,PhysRevLett.132.226001,PhysRevB.98.104505,PhysRevResearch.5.023177}. Proposed mechanisms include 
inter-valley scattering~\cite{PhysRev.134.A511,PhysRev.163.380}, 
coupling to polar~\cite{doi:10.1073/pnas.1604145113}, and defect~\cite{Gorkov2017}
phonons, higher-order electron-phonon interactions~\cite{PhysRevLett.132.226001}, coupling to
the soft infrared active zone-center phonon of $\text{SrTiO}_3$~\cite{PhysRevB.98.104505}, which is enhanced by anharmonic damping~\cite{PhysRevB.105.L020506}, and Rashba coupling~\cite{PhysRevResearch.5.023177,saha2024strongcouplingtheorysuperconductivity}. 
However, 
the emergence of superconductivity through these mechanisms generally depends on parameters that are challenging to compute from first principles or to accurately infer from experiment~\cite{annurev:/content/journals/10.1146/annurev-conmatphys-031218-013144,GASTIASORO2020168107}. A fully first-principles 
explanation for the phonon-mediated electron-electron attraction remains elusive, as 
$\text{SrTiO}_3$ 
becomes
superconducting at low doping densities where
the Fermi energy is much smaller than the phonon frequencies, placing it in the non-adiabatic limit.
As a result, 
the widely used Migdal-Eliashberg framework for phonon-mediated superconductivity does
not apply~\cite{PhysRevB.87.024505},  
leading
to an ongoing debate on the
mechanism of Cooper pair formation.

Here we present a theory that rigorously incorporates lattice screening in electron-electron interactions. Without relying on adjustable parameters, we analytically demonstrate that lattice screening from multiple longitudinal optical (LO) phonons yields long-range attraction and superconductivity in $\text{SrTiO}_3$, in good agreement with experiments. 
Moreover, we translate our theory to a first-principles computational scheme, and show that
the deformation potential associated with vibrations of the Ti-O bond
provides an additional short-range
attractive contribution to electron-electron interactions.
Our theory, outlined in detail in our companion paper~\cite{joint_article}, applies to several doped semiconductors
and insulators. 

The density-density interaction between electronic states $i,j$ is
\begin{align}
    \label{eq:coulomb}
    U_{i\mathbf{R}j\mathbf{R}'}(\omega)=\nonumber \\ \int_V d\mathbf{r} \int_V d\mathbf{r}' \phi_{i\mathbf{R}}^*(\mathbf{r})\phi_{i\mathbf{R}}(\mathbf{r})W(\mathbf{r},\mathbf{r}',\omega)\phi_{j\mathbf{R}'}^*(\mathbf{r}')\phi_{j\mathbf{R}'}(\mathbf{r}').
\end{align}
Here $\phi_{i\mathbf{R}}(\mathbf{r})$ are a
basis localized in real space, specifically maximally localized Wannier functions (MLWFs)~\cite{RevModPhys.84.1419} centered around lattice vectors $\mathbf{R},\mathbf{R}'$. Additionally, $W(\omega)$ is the frequency-dependent screened Coulomb interaction, which has two contributions: an electronically screened Coulomb interaction $W^{el}$, and a contribution from 
phonons. Within many-body perturbation theory and to the lowest order in the electron-phonon interaction, the phonon contribution to the screened Coulomb interaction is written as~\cite{Baym1961,Hedin1965,Giustino2017}
\begin{equation}
\label{eq:phonon_screening}
W^{ph}(\mathbf{r},\mathbf{r}',\omega) = \sum_{\mathbf{q},\nu} D_{\mathbf{q},\nu}(\omega)g_{\mathbf{q},\nu}(\mathbf{r})g^*_{\mathbf{q},\nu}(\mathbf{r}').
\end{equation}
Here $D_{\mathbf{q},\nu}(\omega)$ is the propagator of a phonon with branch index $\nu$ at wavevector $\mathbf{q}$, and $g_{\mathbf{q},\nu}$ is the electron-phonon
vertex. By inserting the overall screened Coulomb interaction $W(\omega)=W^{el}(\omega)+W^{ph}(\omega)$ into eq.\,\eqref{eq:coulomb}, we obtain the total interaction for electrons in states $\phi_{i\mathbf{R}},\phi_{j\mathbf{R}'}$ 
as $U_{i\mathbf{R}j\mathbf{R}'}(\omega)=U^{el}_{i\mathbf{R}j\mathbf{R}'}(\omega)+U^{ph}_{i\mathbf{R}j\mathbf{R}'}(\omega)$.

For the phonon-mediated interaction, as detailed in our
companion paper~\cite{joint_article}, we obtain for periodic systems
\begin{align}
    \label{eq:Uph}
    U^{ph}_{i\mathbf{R}j\mathbf{R}'}(\omega)=\bra{\phi_{i\mathbf{R}}\phi_{j\mathbf{R}'}}W^{ph}(\omega)\ket{\phi_{i\mathbf{R}}\phi_{j\mathbf{R}'}} \nonumber\\ =\sum_{\mathbf{q},\nu}g_{ii\mathbf{q}\nu}(\mathbf{0})g^*_{jj\mathbf{q}\nu}(\mathbf{R}'-\mathbf{R})\nonumber \\
    \times [\frac{1}{\omega-\omega_{\mathbf{q},\nu}+i\delta}-\frac{1}{\omega+\omega_{\mathbf{q},\nu}-i\delta}],
\end{align}
where $g_{mn\mathbf{q}\nu}(\mathbf{R}) = \bra{\phi_{m\mathbf{R}}}g_{\mathbf{q}\nu}\ket{\phi_{n\mathbf{R}}}$. Here we ignore non-local couplings of the form $\bra{\phi_{m\mathbf{R}}}g_{\mathbf{q}\nu}\ket{\phi_{n\mathbf{R}'}}$, which are expected to be small in the Wannier representation.
In the static limit, and for on-site ($\mathbf{R}=\mathbf{R}'$) intra-band ($i=j$) terms
this becomes
\begin{equation}
    \label{eq:Uph_st_on_site}
    U^{ph,st}_{i\mathbf{R}i\mathbf{R}}=-2\sum_{\mathbf{q}\nu}\frac{|g_{ii\mathbf{q\nu}}(\mathbf{0})|^2}{\omega_{\mathbf{q},\nu}},
\end{equation}
which is a purely attractive interaction, in agreement
with the conventional understanding of the effect of phonon screening on electron-electron interactions~\cite{MahanGeraldD2013MP}. The superscript ``$st$" denotes the static limit $\omega=0$.

In polar semiconductors and insulators, linear long-range dipolar electron-phonon coupling 
dominates the 
electron-phonon interaction~\cite{PhysRevB.13.694}, and is written as
\begin{align}
\label{eq:generalized_Frohlich}
    g^{\mathcal{L}}_{ij\mathbf{q}\nu}(\mathbf{R})=i\frac{4\pi}{V}\sum_{\kappa}(\frac{1}{2NM_{\kappa}\omega_{\mathbf{q},\nu}})^{1/2}\frac{\mathbf{q}\cdot \mathbf{Z}_{\kappa}\cdot \mathbf{e}_{\kappa \nu}}{\mathbf{q}\cdot \mathbf{\epsilon}_{\infty}\cdot \mathbf{q}} \nonumber \\ \times \bra{\phi_{i\mathbf{R}}(\mathbf{r})}e^{i\mathbf{q}\cdot \mathbf{r}}\ket{\phi_{j\mathbf{R}}(\mathbf{r})},
\end{align}
which is the generalization of the Fr\"ohlich model~\cite{Verdi2015}. Here we ignore local field effects, \emph{i.e.} we set $\mathbf{q}+\mathbf{G}\rightarrow \mathbf{q}$, where $\mathbf{G}$ a reciprocal lattice vector. 
For an atom $\kappa$, $M_{\kappa}$ and $\mathbf{Z}_{\kappa}$ are its
mass and Born effective charge tensor respectively, $\mathbf{e}_{\kappa \nu}$ a vibrational eigenvector, and
$\mathbf{\epsilon}_{\infty}$ the dielectric matrix tensor. 

\begin{figure*}[tb]
    \centering
    \includegraphics[width=0.8\linewidth]{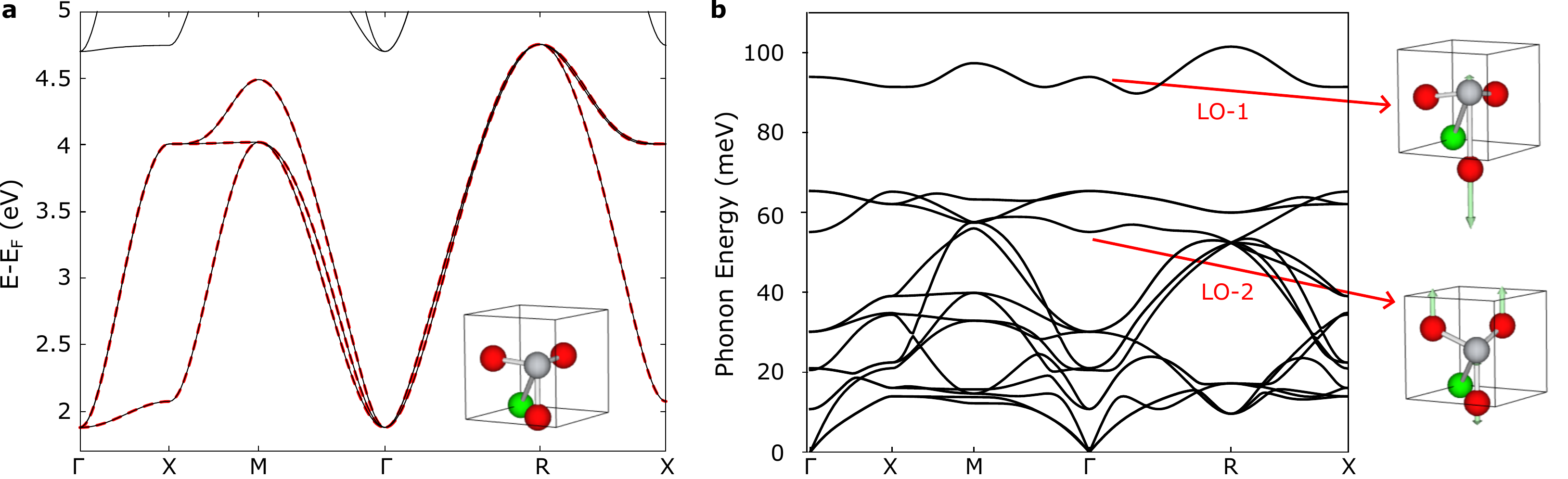}
    \caption{Lowest conduction band manifold of $\text{SrTiO}_3$ computed at the DFT-LDA level (black) and using Wannier interpolation (red) in panel \textbf{a} (energies relative to valence band maximum). The inset visualizes the unit cell of the system (Sr: green, Ti: gray, O: red). Panel \textbf{b} shows the phonon dispersion as obtained within DFPT including anharmonic corrections, alongside a visualization of the displacement patterns of the two highest frequency LO phonons.}
    \label{fig:STO_bands}
\end{figure*}

\begin{table*}[tb]
\centering
  \setlength{\tabcolsep}{6pt} 
\begin{tabular}{ccccccccc}
\hline
$\epsilon_{\infty}$ & $\epsilon_0$ & $Z^*_{\text{Sr}}$ & $Z^*_{\text{Ti}}$  & $Z^{1*}_{\text{O}}$  & $Z^{2*}_{\text{O}}$ & $\omega_{TO-1},\omega_{LO-1}$ (meV) & $\omega_{TO-2},\omega_{LO-2}$ (meV) & $\omega_{TO-3},\omega_{LO-3}$ (meV)\\
\hline
$6.2$ & $409$ & $2.56$ & $7.25$ & $-5.69$ & $-2.05$ & $65,94$ & $21,55$ & $11,21$ \\
\hline
\end{tabular}
\caption{DFPT results for SrTiO$_3$: high-and low-frequency dielectric constants $\epsilon_{\infty},\epsilon_0$, Born effective charges of different atoms (along two directions for oxygen), and frequencies of the pairs of TO,LO modes at $\Gamma$, for the three IR active LO phonons.}
\label{table:STO_properties}
\end{table*}

By using the expression of eq.\,\eqref{eq:generalized_Frohlich} in the phonon-mediated electron-electron interaction of eq.\,\eqref{eq:Uph}, taking the static limit, and assuming an isotropic dielectric and dispersionless phonons, 
we obtain the expression
\begin{align}
    \label{eq:Uph_long_range_static}
    U^{ph,st,\mathcal{L}}_{\mathbf{R}\mathbf{R}'}= -\frac{\gamma^2}{\epsilon_{\infty}|\mathbf{R}-\mathbf{R}'|},
\end{align}
where 
$\gamma^2=\sum_{m}\frac{\omega_{LO,m}^2-\omega_{TO,m}^2}{\omega_{LO,m}^2}$,
and $m$ runs over LO-TO phonon pairs of the system.
This is an attractive interaction, with the contribution of each LO mode being proportional to
the splitting $\omega_{LO,m}^2-\omega_{TO,m}^2$ from the corresponding transverse optical (TO) mode. This result is derived within our companion paper~\cite{joint_article}, alongside expressions for the general frequency-dependent case, which accounts
for anisotropic dielectric and dispersive phonons. By comparing this expression to a repulsive electron-electron interaction $1/(\epsilon_{\infty}|\mathbf{R}-\mathbf{R}'|)$, it is evident that an overall long-range attraction
appears when $\gamma^2>1$.

Superconductivity in $\text{SrTiO}_3$ emerges upon doping of the conduction bands, it is therefore critical to incorporate doped carriers into our theoretical framework in order to gain microscopic insights into this process. At low doping concentrations $n$ below
a critical Mott density $n_c$, doped carriers are localized and the system remains insulating, albeit with a modified dielectric constant. This can be estimated from a simple Lorentz-Lorenz model at the critical density as
$\bar{\epsilon}_{\infty}=\epsilon_{\infty}+4\pi n_c \alpha$~\cite{RevModPhys.53.81},
where the polarizability $\alpha=9(\epsilon_0/m^*)^{3/2}$ for a hydrogen-like defect. We compute using DFT/LDA and a $GW$ correction of the band structure an electron effective mass of $m^*=0.39$. The critical density is obtained via the Mott criterion $n_c=0.26(\frac{m^*}{\epsilon_0})^3=1.5\times10^{15}$$\text{cm}^{-3}$, in reasonable agreement with the experimental value of $2\times10^{16}$\,$\text{cm}^{-3}$~\cite{PhysRevB.81.155110}. Beyond this critical density, an insulator to metal transition occurs, and doped carriers behave as free particles described by a Thomas-Fermi screening model. We thus generalize 
our expression of eq.\,\eqref{eq:Uph_long_range_static} for the phonon-mediated electron-electron interaction to include screening from localized and free carriers, and in reciprocal space
the total interaction becomes~\cite{Ren2020,doi:10.1073/pnas.1604145113}:
\begin{align}
       \label{eq:final_Coulomb} U^{st,\mathcal{L}}_{\mathbf{k}\mathbf{k'}}
        =\frac{4\pi}{\bar{\epsilon}_{\infty}}\cdot\frac{1-\gamma^2}{|\mathbf{k-k'}|^2+ \kappa^{2}_{TF}} ,
\end{align}
where the Thomas-Fermi screening length
$\kappa^{2}_{TF} = \frac{4m^*p_{F}}{\bar{\epsilon}_{\infty}\pi}$,
with the doping-dependent Fermi momentum $p_F$ determined from the computed density of states. A more rigorous treatment of electron-phonon coupling incorporating doping is possible within a first-principles context~\cite{PhysRevLett.129.185902}. 

For a single dominant LO phonon mediating the Coulomb interaction, Gor'kov derived an expression for the superconducting critical temperature in the non-adiabatic limit, where $\text{SrTiO}_3$ lies~\cite{doi:10.1073/pnas.1604145113}. In Appendix\,\ref{T_c} we generalize this using
our expression for the Coulomb interaction incorporating long-range coupling to multiple longitudinal optical phonons and we obtain:
\begin{align}
\label{eq:Tc}
        T_c = \frac{m^*\gamma}{\pi^3\bar{\epsilon}_{\infty}^2}x^2\exp(-\frac{2x}{(\gamma^2-1)\ln(1+x)}),
\end{align}
where $x=\pi p_F\bar{\epsilon}_{\infty}/m^*$.

Above $105$\,K,  $\text{SrTiO}_3$ assumes
a high-symmetry cubic perovskite structure, with perfectly aligned TiO$_6$ octahedra. The unit cell
in this phase, shown in Fig.\,\ref{fig:STO_bands}a, 
consists of five atoms. Below $105$\,K the system is
in its tetragonal paraelectric phase~\cite{COWLEY1969181,doi:10.1143/JPSJ.26.396}, where
the TiO$_6$ octahedra undergo tilting,   doubling the number of atoms in the unit cell. Here we focus on cubic $\text{SrTiO}_3$ due to the qualitatively similar but simpler character of its phonons and electron-phonon coupling relative to the tetragonal phase. Fig.\,\ref{fig:STO_bands}a shows the first three conduction bands of $\text{SrTiO}_3$, as obtained within density functional theory (DFT) and the local density approximation (LDA). 
Fig.\,\ref{fig:STO_bands}b visualizes the $\text{SrTiO}_3$ phonon dispersion, in excellent agreement with previous calculations~\cite{PhysRevLett.121.226603,PhysRevB.108.035155}, as computed within 
density functional perturbation theory (DFPT), corrected for anharmonicity using finite differences, as outlined in Appendix\,\ref{computational_details}. The anharmonic correction is important, as for $\text{SrTiO}_3$, harmonic phonons computed at the LDA level of theory show instabilities at the $R$ and $M$ points in the Brillouin zone~\cite{Zhang2017}.  
Table\,\ref{table:STO_properties} summarizes the DFPT
results for the dielectric constants, Born effective charges, and LO/TO mode frequencies of $\text{SrTiO}_3$. Cubic $\text{SrTiO}_3$ has four LO phonons, three of which are IR active, and the frequencies of these are given in Table\,\ref{table:STO_properties}. Additionally, 
Fig.\,\ref{fig:STO_bands}b illustrates the displacement patterns of the two highest frequency LO modes,
which involve vibrations of the Ti-O bond. In Appendix\,\ref{computational_details} we provide all computational details.

\begin{figure}[tb]
    \centering
    \includegraphics[width=0.8\linewidth]{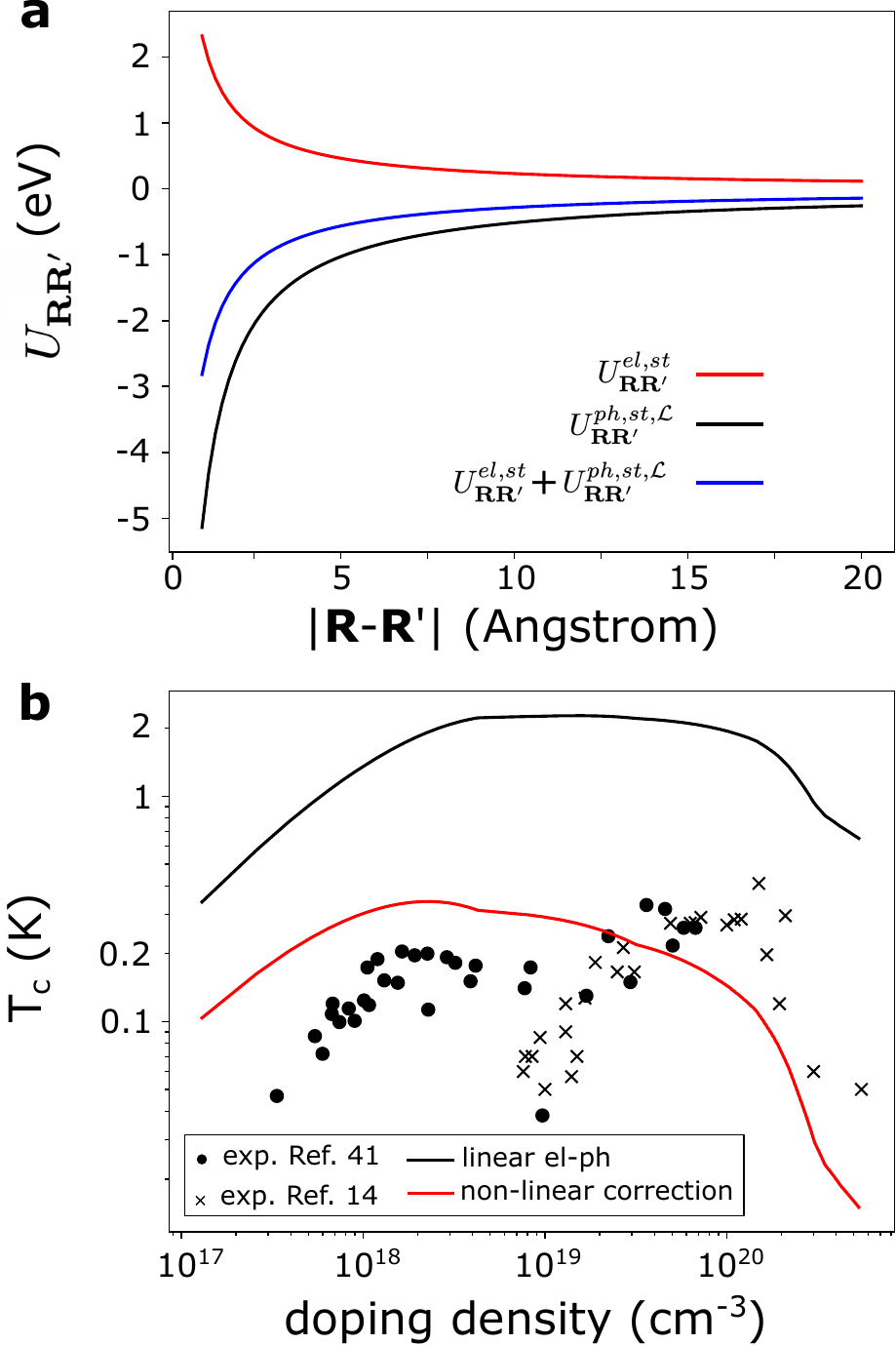}
    \caption{Electron-electron Coulomb repulsion, phonon-mediated long-range electron attraction, and total interaction (panel \textbf{a}) as a function of distance between electrons. In panel \textbf{b} we visualize the superconducting critical temperature of $\text{SrTiO}_3$ according to our eq.\,\eqref{eq:Tc} (black), and also for including a correction for non-linear electron-phonon coupling (red). Experimental values are taken from Refs.~\cite{PhysRev.163.380,Rischau2017}.   }
    \label{fig:STO_vs_R}
\end{figure}

We now visualize in Fig.\,\ref{fig:STO_vs_R}a the static long-range phonon-mediated
electron-electron interaction of cubic $\text{SrTiO}_3$ in the undoped case, according to
equation\,\eqref{eq:Uph_long_range_static}, and we compare to the long-range electronic repulsion $U^{el}=1/(\epsilon_{\infty}|\mathbf{R}-\mathbf{R}'|)$. The phonon-mediated attraction (black) clearly overcomes the Coulomb
repulsion (red) between electrons, resulting in a net attraction (blue). This results in the formation of Cooper pairs upon doping the conduction bands, and the critical temperature of superconductivity is visualized as a function of doping density in Figure\,\ref{fig:STO_vs_R}b (black curve). We see that we reproduce the experimental dome-like behavior semi-quantitatively (experimental data are obtained from Refs.~\cite{PhysRev.163.380,Rischau2017}), although we generally overestimate the experimental $T_c$, and our predicted critical temperature is only accurate up to a factor of order one, as outlined in Appendix\,\ref{T_c}. 
It is worth emphasizing that this result has been obtained by accounting for the linear long-range coupling to all LO phonons rigorously and on equal footing,
and approximately incorporating doping in our Coulomb interaction of eq.\,\eqref{eq:final_Coulomb}. However, higher-order multi-phonon effects have been discussed to be important in $\text{SrTiO}_3$~\cite{PhysRevB.102.045126,PhysRevB.108.035155}, and specifically in the context of superconductivity~\cite{PhysRevResearch.1.013003,PhysRevB.104.L220506,Volkov2022}. In Appendix\,\ref{multi_phonon} we employ finite differences to estimate that on average, these effects can indeed be significant at higher temperatures, while at temperatures near $0$\,K they provide an effective reduction of $g^2$ to $0.83$ of its
value when only accounting for linear electron-phonon coupling.
The red curve in Figure\,\ref{fig:STO_vs_R}b accounts
for such a rescaling $g^2\rightarrow0.83g^2$ from multi-phonon effects, bringing the predicted critical temperatures closer to experiment. Nevertheless, the qualitative behavior of the critical temperature, as a result of the linear coupling of electrons to multiple LO phonons, remains unchanged. 


While the attractive interaction due to long-range electron-phonon coupling is sufficient to give rise to superconductivity, other mechanisms have also
been discussed as potentially playing a role, such as the deformation potential mechanism~\cite{PhysRevB.98.104505,PhysRevB.100.226501}. In order to evaluate the relevance of such mechanisms to electron pairing in $\text{SrTiO}_3$, and
due to the lack of analytic expressions for such electron-phonon interactions, we return to eq.\,\eqref{eq:Uph} and evaluate this
term from first principles including only short-range electron-phonon coupling $g^{\mathcal{S}}_{ii\mathbf{q}\nu}(\mathbf{R})$. Details of our first-principles implementation of eq.\,\eqref{eq:Uph} are given in our companion paper~\cite{joint_article}. 
Briefly, we use Quantum Espresso for DFT calculations~\cite{QE}, and Wannier90~\cite{Pizzi2020} to produce MLWFs used as a basis for the
terms in eqs.\,\eqref{eq:coulomb} and\,\eqref{eq:Uph}. The resulting MLWFs
describing the conduction states of $\text{SrTiO}_3$ resemble Ti $d$ orbitals, and yield a Wannier-interpolated band structure in excellent agreement with DFT (Fig.\,\ref{fig:STO_bands}). We use RESPACK to evaluate
the electronic Coulomb repulsion according to eq.\,\eqref{eq:coulomb} for the subspace of the
three conduction states, within the constrained random phase approximation (cRPA)~\cite{PhysRevB.70.195104}. We use EPW~\cite{Lee2023}, BerkeleyGW~\cite{Deslippe2012} and Abinit~\cite{GONZE2020107042}, allowing us to to compute the 
phonon-mediated electron interaction following eq.\,\eqref{eq:Uph} within $GW$ perturbation theory (GWPT)~\cite{PhysRevLett.122.186402}, which captures the effect of many-electron correlations on the electron-phonon coupling.  

\begin{figure}[tb]
    \centering
    \includegraphics[width=0.7\linewidth]{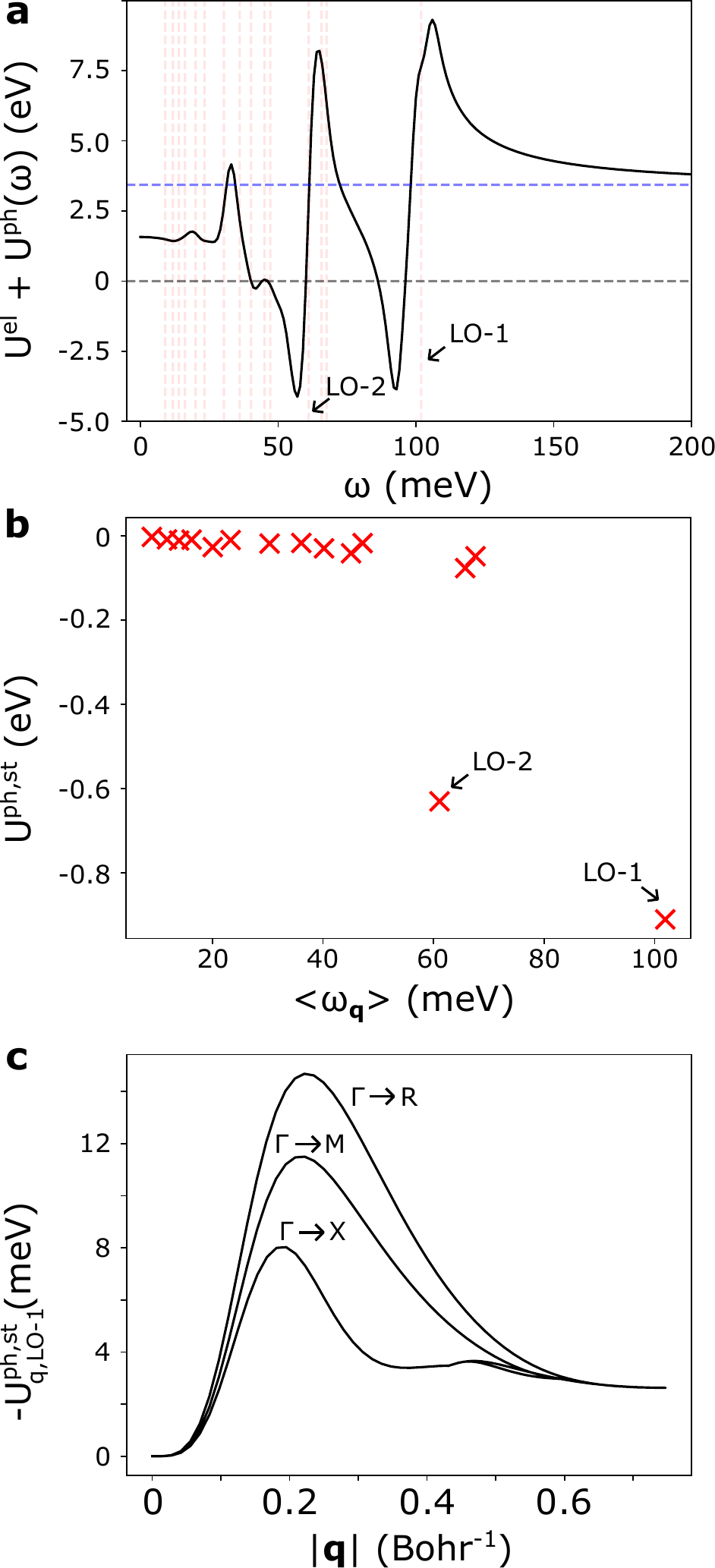}
    \caption{On-site electron-electron interaction in $\text{SrTiO}_3$, mediated by short-range coupling to phonons. In panel \textbf{a} we plot the frequency dependence of the total interaction, where red lines are average phonon frequencies in the Brillouin zone, and the blue dashed line marks the static purely electronic interaction. In panel \textbf{b} we decompose the static on-site attraction to the contributions of different phonons, and panel \textbf{c} shows how the contribution of LO-1 depends on its momentum $|\mathbf{q}|$.}
    \label{fig:STO_modes}
\end{figure}

The most important contribution of the phonon-mediated attraction due to short-range electron-phonon coupling is
to the on-site ($\mathbf{R}=\mathbf{R}'$) Coulomb term, where an attraction of $U^{ph,st,\mathcal{S}}_{\mathbf{R}\mathbf{R}}=-1.858$\,eV is found. Within cRPA we find that the on-site repulsion
between two conduction electrons is $U^{el,st}_{\mathbf{R}\mathbf{R}}=3.436$\,eV. Therefore, while there is a substantial reduction of $54\%$ in the on-site
Coulomb repulsion due to phonons, the overall interaction remains repulsive.
To better understand the attractive contribution from short-range electron-phonon
coupling, we visualize in Fig.\,\ref{fig:STO_modes}a the frequency 
dependence of the total on-site electron-electron interaction, including phonon effects. Here we have  applied a Gaussian smoothing
filter to reduce noise. 
We see that there are poles associated with
the LO phonons, and it is only at fairly large
frequencies of approximately $50$\,meV that the
overall on-site interaction becomes attractive. 
While the static long-range phonon-mediated interaction between electrons
is already sufficient to yield superconductivity, it is worth noting that
this on-site interaction, which only becomes attractive at finite frequencies,
could also contribute to superconductivity, as it has been shown in Hubbard-Holstein models with appropriate retardation and on-site interaction strength~\cite{PhysRevLett.125.167001}.
In Fig.\,\ref{fig:STO_modes}b we visualize
the contribution of individual phonons to
the on-site phonon-mediated attraction in
the static limit. This effect
is dominated by the two highest frequency LO
modes and we decompose the contribution of
the highest frequency one  (LO-1) along three high-symmetry paths
in the Brillouin zone in Fig.\,\ref{fig:STO_modes}c. We see that the absolute value of the phonon-mediated attraction initially grows with $|\mathbf{q}|$,
and eventually peaks. This is consistent with analytic
theories of electron-phonon coupling~\cite{PhysRev.183.730,PhysRevB.13.694}, where the
deformation potential includes octopole and higher-order terms, which vanish as $\mathbf{q}\rightarrow 0$.


It is worth asking what makes $\text{SrTiO}_3$ special in terms of the emergence of attractive electron-electron interactions, and
whether similar behavior might be observed in other materials. As presented above, there are two mechanisms
for phonon-mediated attraction at play. The
first is the long-range coupling of electrons
to multiple LO phonons. As expressed by eq.\,\eqref{eq:Uph_long_range_static}, this channel
is significant in materials where multiple LO 
phonons have a substantial splitting from the corresponding TO phonons - a manifestation of large Born effective charges. The
second contribution to a strong phonon-mediated electron attraction is the deformation potential mechanism, which is generally associated with strong
covalent character of the bonds of a material~\cite{Cohen_Louie_2016}. As seen
in Table\,\ref{table:STO_properties} the Born effective charges of the Ti and O values are much greater than 
their nominal charges, reflecting the ionic-covalent character of the Ti-O bond, similar to other cubic perovskites~\cite{PhysRevLett.72.3618}. A similar phonon-mediated attraction due to a deformation potential mechanism was found for GeTe in our companion paper~\cite{joint_article}. Our results are consistent with
the anomalous isotope effect of $\text{SrTiO}_3$, where superconductivity is enhanced upon substitution of $^{16}O$ by $^{18}O$~\cite{Stucky2016}. Oxygen substitution 
reduces the frequency of the lowest TO mode~\cite{PhysRevLett.115.247002}, which in turn
will increase the attraction between electrons
mediated by the deformation potential mechanism, given the $1/\omega_{\mathbf{q},\nu}$ dependence of the phonon-mediated attraction $U^{ph}$. 

We have presented a theoretical framework for the phonon-mediated electron interactions of materials, and applied it to $\text{SrTiO}_3$. We have shown that
long-range coupling of electrons to multiple LO phonons
can result in superconductivity, in good agreement with experiments. Moreover, 
we have derived a simple criterion to determine
whether this mechanism appears within a given material. It will be interesting moving forward to examine in more detail the importance of long-range electron-phonon coupling for inducing superconductivity, as has been previously discussed in the context of high-temperature superconductivity~\cite{annurev:/content/journals/10.1146/annurev-conmatphys-033117-053942}. 
Furthermore, we have combined our theory with
first-principles calculations, and found
a strong short-range phonon-mediated attraction of electrons, following a deformation potential mechanism, which is expected to be strong in systems with mixed ionic-covalent bonding. We expect that our theoretical and computational framework will contribute towards a deeper
understanding of phonon-mediated superconductivity in
a wide range of materials. 

The authors are thankful to Marvin L. Cohen for
helpful and inspiring discussions. 
This material is based upon work supported by the U.S. Department of Energy, Office of Science, National Quantum Information Science Research Centers, Superconducting Quantum Materials and Systems Center (SQMS) under contract No. DE-AC02-07CH11359.  We are grateful for support from NASA Ames Research Center.  C. J. N. C. and M. R. F. acknowledge support from the UK Engineering and Physical Sciences Research Council (EPSRC). VV was supported by the National Science Foundation (NSF) CAREER award through grant No. DMR-1945098. This research used resources of the National Energy Research
Scientific Computing Center, a DOE Office of Science User Facility supported by the Office of Science of the U.S. Department of Energy under Contract No. DE-AC02-05CH11231 using NERSC awards HEP-ERCAP0029167 and DDR-ERCAP0029710.

\clearpage

\appendix

\section{Computational details}
\label{computational_details}

We employ Quantum Espresso~\cite{QE} for DFT and DFPT calculations, Wannier90~\cite{Pizzi2020} to generate Wannier
states, and EPW~\cite{Lee2023} in
order to perform Wannier-Fourier interpolation of electron-phonon matrix elements. 
We use scalar-relativistic optimized norm-conserving Vanderbilt pseudopotentials (ONCV)~\cite{Hamann2013} with standard accuracy, taken from Pseudo Dojo~\cite{VanSetten2018}. 
We work within the local density approximation (LDA) of DFT. However, acoustic
phonons obtained with DFPT become unstable at the $R$ and $M$ points of the Brillouin zone due to anharmonicity. To account for this anharmonicity, 
we displace the structure along an unstable harmonic phonon to find a new
minimum in a supercell structure, and we re-compute the now stable phonons using finite differences~\cite{monserrat2018electron}, in a $2\times 2\times 2$ supercell. For all DFT calculations we use a plane
wave cutoff of $90$\,Ry. For cRPA calculations we
compute the electronic density on a $6\times 6\times 6$ $\Gamma$-centered $\mathbf{k}$-grid, and we include $600$
bands, using a cutoff of $20$\,Ry, and having excluded the three
lowest-lying conduction bands. For DFPT and GWPT calculations we use an electronic density computed on a $6\times 6\times 6$ half-shifted $\mathbf{k}$-grid, and we obtain the
phonons on a $6\times 6\times 6$ grid of $\mathbf{q}$-points ($3\times 3\times 3$ for both grids for the $2\times 2\times 2$ supercells). We interpolate the phonon frequencies and electron-phonon matrix elements on a $20\times 20\times 20$ $\mathbf{q}$-grid, which converges
the short-range phonon-mediated attractive contribution
to the Coulomb interactions within the conduction bands. It is worth highlighting that GWPT calculations
are known to yield more accurate, and often stronger electron-phonon interactions compared to DFPT~\cite{PhysRevLett.122.186402}. Indeed we
found for $\text{SrTiO}_3$ that using GWPT
increases the absolute value of the short-range on-site correction to the Coulomb interaction due to
phonons by $18\%$, compared to DFPT. 

\section{Cooper instability in Logarithmic approximation}
\label{T_c}

SrTiO$_3$ becomes superconducting at very low doping densities, where $\frac{\omega_{\nu}
}{E_{F}}\gg 1$, with $\omega_{\nu}$ a typical phonon frequency. Following Gor'kov~\cite{PhysRevB.93.054517,doi:10.1073/pnas.1604145113}, we take the logarithmic approach which is applicable in this non-adiabatic limit. Here superconductivity manifests itself in the emergence of a pole in the scattering amplitude of the dynamical screened Coulomb interaction as a result of the Dyson equation
\begin{align}
        \Bar{U}_{\mathbf{k}\mathbf{k'}}(i\omega_n) = \tilde{U}_{\mathbf{k}\mathbf{k'}}(i\omega_n) - \frac{T}{(2\pi)^3}\nonumber \\ \times \sum_{m}\int d\mathbf{k}_1 \tilde{U}_{\mathbf{k}\mathbf{k}_1}(i\omega_n-i\omega_{m})\nonumber \\ \times G_{\mathbf{k}_1}(i\omega_{m})G_{-\mathbf{k}_1}(-i\omega_{m})\Bar{U}_{\mathbf{k}_1\mathbf{k'}}(i\omega_{m}),
\end{align}
where $G$ the electron's Green's function. 
We obtain $T_c$ from the eigenvalue of the homogeneous equation for the gap function $\Delta$:
\begin{align}
        \Delta_{\mathbf{k}}(i\omega_n) = -T\sum_{m}\int d\mathbf{k}_1 \tilde{U}_{\mathbf{k}\mathbf{k}_1}(i\omega_n-i\omega_{m})\nonumber \\ \times \Pi_{\mathbf{k}_1}(i\omega_{m})\Delta_{\mathbf{k}_1}(i\omega_{m}),
\end{align}
with the RPA polarization
\begin{align}
        \Pi_{\mathbf{k}}(i\omega_{m}) = G_{\mathbf{k}}(i\omega_{m})G_{-\mathbf{k}}(-i\omega_{m})= \frac{1}{\omega^2_{m}+\Big(\frac{k^2-k^2_F}{2m^*}\Big)^2},
\end{align}
where we have made a parabolic band approximation. 
Solving for the gap function exactly gives $T_c$ in a BCS weak-coupling form 
\begin{align}
        T_c \propto \exp(-\frac{1}{\lambda}).
\end{align}
We now determine $\lambda$ without solving the gap equation. 
Substituting the RPA polarization into the gap equation, we obtain
\begin{align}
\label{eq:gap_equation}
        \Delta_{\mathbf{k}}(i\omega_n)=-T \sum_{m}\int d\theta d\xi \tilde{U}_{\mathbf{k}\mathbf{k}_1}(i\omega_n-i\omega_m)\frac{m^*p_F\sin\theta}{(2\pi)^2}\nonumber \\ \times \frac{1}{\omega^2_{m}+\xi^2} \Delta_{\mathbf{k}_1}(i\omega_m).
\end{align}
Let $\mathbf{k},\mathbf{k}_1$ be on the Fermi surface, (meaning $|\mathbf{k}|^2=|\mathbf{k}_1|^2=|\mathbf{k}_F|^2$), and
\begin{align}
        \tilde{U}_{\mathbf{k}\mathbf{k}_1}(i\omega_n-i\omega_m) \rightarrow     \tilde{U}_{\mathbf{k}\mathbf{k}_1}(0) =\frac{4\pi}{\bar{\epsilon}_{\infty}}\cdot\frac{1-\gamma^2}{|\mathbf{k}-\mathbf{k}_1|^2+ \kappa^{2}_{TF}}.
\end{align}
Given this form for the Coulomb interaction, the right-hand side of the gap equation eq.\,\eqref{eq:gap_equation} no longer depends on $\omega_n$, meaning that $\Delta_{\mathbf{k}}(i\omega_n)$ must be a constant, and can be taken out of the integral. Therefore, we can devide both sides of the equation by $\Delta_{\mathbf{k}}(i\omega_n)$, and eq.\,\eqref{eq:gap_equation} simplifies to
\begin{align}
        1=\Bigg(\int^{\pi}_0 d\theta \tilde{U}_{\mathbf{k}\mathbf{k}_1}(\omega=0)\frac{m^*p_F\sin\theta}{(2\pi)^2}\Bigg) \nonumber \\ \times \int^{W}_0 \frac{d\xi }{\xi} \tanh\Big(\frac{\xi}{2T_c}\Big),
\end{align}
where $W=A\cdot E_F$ the relevant energy scale, with $A$ a constant of order unity.
This may be rewritten as 
\begin{align}
        1 = \lambda \ln\Big(\frac{2W\gamma}{\pi T_c}\Big),
\end{align}
from where we obtain 
\begin{align}
\label{eq:lambda_Tc}
        T_c = \frac{2W\gamma}{\pi}\exp(-\frac{1}{\lambda}).
\end{align}
The coupling constant $\lambda$ is
\begin{align}
        \lambda = \int^{\pi}_0 d\theta \tilde{U}_{\mathbf{k}\mathbf{k}_1}(\omega=0)\frac{m^*p_F\sin\theta}{(2\pi)^2}\nonumber \\
        = \frac{1-\gamma^2}{2\pi\bar{\epsilon}_{\infty}}\int^{\pi}_0 d\theta \frac{m^*\sin\theta}{p_F(1-\cos\theta+\frac{\kappa^2_{TF}}{2p^2_F})},
\end{align}
and $p^2_F = 2m^* E_F$. We evaluate the integral to obtain
\begin{align}
        \lambda = \frac{m^*(\gamma^2-1)}{2\pi p_F \bar{\epsilon_{\infty}}}\ln\Big(1+\frac{\pi p_F\bar{\epsilon}_{\infty}}{m^*} \Big).
\end{align}
Now, we define the variable $x=\pi p_F \bar{\epsilon}_{\infty}/m^*$ to get 
\begin{align}
         \lambda = \frac{\gamma^2-1}{2x}\ln\Big(1+x\Big).
\end{align}
By using this expression, substituting $W$, using the definition of the variable $x$, and given that $2E_F = \frac{p^2_F}{m^*}$, the critical temperature of eq.\,\eqref{eq:lambda_Tc} becomes
\begin{align}
        T_c = A\frac{m^*\gamma}{\pi^3\bar{\epsilon}_{\infty}^2}x^2\exp(-\frac{2x}{(\gamma^2-1)\ln(1+x)}).
\end{align}
This expression reproduces the characteristic form of $T_c$ as a function of charge carrier density.  


\section{Impact of multi-phonon effects on the electron-phonon interaction}
\label{multi_phonon}

\begin{figure}[b]
    \centering
    \includegraphics[width=\linewidth]{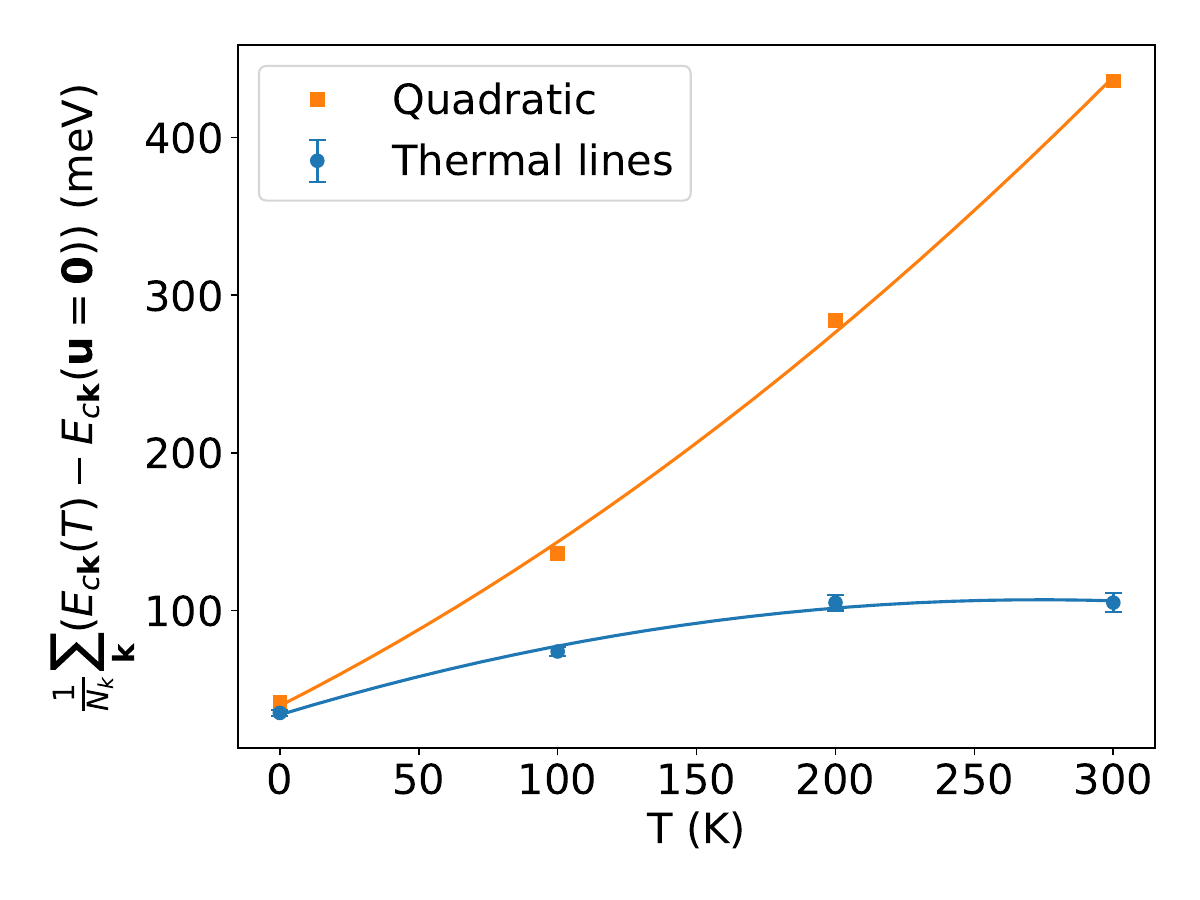}
    \caption{Average renormalization of the $\text{SrTiO}_3$ conduction band energy across the Brillouin zone, as a function of temperature, using the thermal lines and quadratic methods.}
    \label{fig:Ecb_vs_T}
\end{figure}

In order to evaluate the importance of higher-order electron-phonon coupling, 
we evaluate the coupling of conduction electrons to phonons using finite differences~\cite{monserrat2018electron}. The
vibrational average of the energy of the lowest conduction state $E_{c\mathbf{k}}$, at a given $\mathbf{k}$-point and
temperature $T$, may be written as
\begin{equation}
\label{eq:exp_value}
    E_{c\mathbf{k}}(T)=\frac{1}{\mathcal{Z}}\sum_{\mathbf{s}}\bra{\chi_{\mathbf{s}}(\mathbf{u})}E_{c\mathbf{k}}(\mathbf{u})\ket{\chi_{\mathbf{s}}(\mathbf{u})} e^{-E_{\mathbf{s}}/k_{\mathrm{B}}T},
\end{equation}
where $\mathbf{u}$ a displaced nuclear configuration of the system, $\mathcal{Z}$ the partition function, and $\ket{\chi_{\mathbf{s}}(\mathbf{u})}$ a vibrational eigenstate with energy $E_{\mathbf{s}}$. 
Equation\,\eqref{eq:exp_value} includes electron-phonon interactions to all orders. We sample the integral of eq.\,\eqref{eq:exp_value} using $100$ thermal lines~\cite{PhysRevB.93.014302}, which result to a narrow distribution around the average and correspond to displaced configurations based on the anharmonic phonons of $\text{SrTiO}_3$. In Fig.\,\ref{fig:Ecb_vs_T} we visualize the
shift of the lowest energy conduction state compared to the value without phonon effects, averaged over the Brillouin zone, \emph{i.e.}, $\Delta E=\frac{1}{N_k}\sum_{\mathbf{k}}(E_{c\mathbf{k}}(T)-E_{c\mathbf{k}}(\mathbf{u}=\mathbf{0}))$, at $T=0,100,200,300$\,K.

We now expand $E_{c\mathbf{k}}(\mathbf{u})$ in $\mathbf{u}$, truncate to second order, and write
\begin{align}
    \label{eq:quadratic}
    E_{c\mathbf{k}}(T) - E_{c\mathbf{k}}(\mathbf{u}=\mathbf{0})\nonumber \\=\sum_{\mathbf{q},\nu} \frac{1}{2\omega_{\mathbf{q},\nu}}\cdot \frac{\partial^2E_{c\mathbf{k}}}{\partial u_{\mathbf{q},\nu}^2}[\frac{1}{2}+n_B(\omega_{\mathbf{q},\nu},T)].
\end{align}
This quadratic formula is the finite-difference equivalent of the Allen-Heine-Cardona theory for temperature-dependent band structures~\cite{allen1976theory,PhysRevB.23.1495}, which is obtained from
accounting for linear electron-phonon coupling, and where the correction is proportional to $g^2$ ($g$ is the electron-phonon matrix element). Therefore, differences in the values of $E_{c\mathbf{k}}(T)$ when using eq.\,\eqref{eq:exp_value} and eq.\,\eqref{eq:quadratic} are due to higher-order
coupling of conduction electrons to phonons, and in Fig.\,\ref{fig:Ecb_vs_T} we also plot the average correction 
to the conduction state energy across the Brillouin zone, as a function of temperature.

We find that while near $T=0$\,K
the two levels of theory are in good agreement, differences become more pronounced with increasing temperature, which is due to the more significant
role of multi-phonon terms, as has been discussed in the literature~\cite{PhysRevB.102.045126,PhysRevB.108.035155}. Nevertheless, the $T=0$\,K limit is the relevant one for superconductivity in $\text{SrTiO}_3$, and in this case we find that 
multi-phonon effects 
reduce 
the phonon-induced renormalization 
to $83\%$ of its value when only considering linear coupling, across the Brillouin zone. We therefore estimate  a similar reduction in the phonon-mediated attraction $U^{ph}\propto g^2/\omega_{\nu}$, once multi-phonon effects near $T=0$\,K are accounted for, and as a correction we rescale $g^2\rightarrow 0.83g^2$. 

\end{document}